\documentstyle[aps,floats,graphicx]{revtex}

\begin{document}

\newcommand{\bec}{\begin{center}}
\newcommand{\ec}{\end{center}}
\newcommand{\be}{\begin{equation}}
\newcommand{\ee}{\end{equation}}
\newcommand{\beqn}{\begin{eqnarray}}
\newcommand{\eeqn}{\end{eqnarray}}
\newcommand{\bet}{\begin{table}}
\newcommand{\ent}{\end{table}}
\newcommand{\bib}{\bibitem}

\wideabs{

\title{
Asymmetric intermixing in Pt/Ti
}

\author{P. S\"ule, M. Menyh\'ard, L. K\'otis, J. L\'ab\'ar, W. F. Egelhoff Jr.$^{\star}$} 
  \address{Research Institute for Technical Physics and Material Science,\\
Konkoly Thege u. 29-33, Budapest, Hungary,sule@mfa.kfki.hu,www.mfa.kfki.hu/$\sim$sule,\\
$^{\star}$ National Institute of Standards \& Technology, Gaithersburg, Maryland 20899\\
}

\date{\today}

\maketitle

\begin{abstract}
The ion-sputtering induced intermixing is studied by Monte-Carlo TRIM, molecular dynamics (MD) simulations,
and Auger electron spectroscopy depth profiling (AES-DP) analysis in Pt/Ti/Si substrate (Pt/Ti) and Ta/Ti/Pt/Si substrate (Ti/Pt) multilayers.
Experimental evidence is found for
the asymmetry of intermixing in Pt/Ti, and in Ti/Pt. 
In Ti/Pt we get a much weaker interdiffusion than in Pt/Ti.
The unexpected enhancement of the interdiffusion of the Pt atoms into the Ti substrate 
has also been demonstrated by simulations.
We are able to capture the essential features of intermixing using TRIM and MD simulations for ion-beam sputtering
and get reasonable values for interface broadening which can be compared with the
experimental measurements.
However, the origin of the asymmetry remains poorly understood
yet.

\end{abstract}
}

\section{Introduction}

 Bombardment of surfaces by energetic particles often leads to atomic inter-layer mixing
\cite{Gnaser,Averback,Ghose,Som,Enrique,Zhou}.
The ion-induced inter-layer atomic transport through interfaces (ion-intermixing) has been the subject of
numerous studies \cite{Averback,Egelhoff,Cai}.
However, recent results indicate that our knowledge on the mechanism of ion-mixing
might be incomplete \cite{Sule_PRB05}.

  Ion-sputtering of surfaces by energetic particles has widely been used for depth profiling
of interfacial structures \cite{sputtering2}.
 Recent advances in the use of ion-sputtering has been attracted considerable
theoretical and experimental attention \cite{sputtering1}.
In particular, ion-sputtering has been simulated by molecular dynamics
using sequential bombardment of metallic surfaces \cite{Karolewski,Thijsse,Zhong,Hanson}.
  In order to get more insight into processes which govern interdiffusion
during repeated ion-bombardments
Auger electron spectroscopy (AES) depth profiling analysis (see e.g. the review refs.  of \cite{Hofmann}) has been used for the
study of
broadening
of interfaces during ion-sputtering in combination with transmission electron
microscopy (TEM) \cite{Gnaser,Menyhard}.

 There is now also a widespread interest in understanding the asymmetry and the
anomalously wide range of intermixing in various diffusion couples \cite{Buchanan,Luo}.
It has also been reported that
interdiffusion is not driven by bulk diffusion parameters nor by
thermodynamic forces (such as heats of alloying) \cite{Sule_NIMB04,Buchanan}.
 Computer simulations have also revealed that
mass-anisotropy of the bilayer governs ion-bombardment induced interdiffusion at the interface \cite{Sule_PRB05} and
greatly influences surface morphology development during ion-sputtering
\cite{Sule_SUCI,Sule_NIMB04_2}.
We also pointed out that interdiffusion takes place via ballistic jumps (ballistic mixing)
in various diffusion
couples under the effect of ion-bombardment \cite{Sule_NIMB04}.

 We demonstrate in the present communication that using atomistic simulations
we are able to reproduce the experimental interfacial broadening and the asymmetry of intermixing obtained by ion-sputtering.
Furthermore, we would like to model and explain the enhancement of intermixing as a function of the anisotropy of the
interfaces.
We reproduce and partly explain the experimentally found asymmetry of intermixing by TRIM and MD simulations.

\section{The setup of the measurements and the simulations}

\subsection{The experimental setup}

  The experimental setup is the following:
According to the crossectional TEM (XTEM) results the thickness of the layers in
samples are:
Pt $13$ nm/Ti $11$ 
nm/Si substrate
(denoted throughout the paper as Pt/Ti), and Ta 21 nm (cap layer to prevent oxidation
of Ti)/Ti 11 nm/Pt 12 nm/Si substrate (denoted as Ti/Pt). 
The XTEM images are shown in Fig 1.
For the sake of simplicity we consider our multilayer samples as bilayers and we study the 
atomic transport processes at the Ti/Pt and Pt/Ti interfaces.
Both samples have been AES depth profiled by applying various sputtering conditions.
The sample has been rotated during sputtering. In the following we will outline results of $500$ eV
 Ar$^+$ ion bombardment at an
angle of incidence of $10^{\circ}$ (with respect to the surface).
The atomic concentrations of Pt, Ti, Ta and Si were calculated by the relative sensitivity method
taking the pure material's values from the spectra. The oxygen atomic concentration has been
calculated by normalizing the measured oxygen Auger peak-to-peak amplitude to TiO$_2$ 
\cite{Vergara}. The depth scale was determined by assuming that the sputtering yield ($Y_i$) in the
mixed layer is the weighted sum of the elemental sputtering yields ($\sum_i X_i Y_i$).
The broadening of the interface is frequently characterizied by the depth resolution. The depth resolution is defined  as the distance of points on the depth profile exhibiting 84 \% and  16 \% concentrations. This definiton has been introduced for the "normal" cases, when either the ion mixing or the roughening can be described by a Gaussian convolution resulting 
in an erf fuction transition in the depth profile. The same definition used for other cases as well, however.

  If the transition does differ from the $erf$ function (e.g. when the mobility of one of the components of a diffusion couple is much higher than that of the other's) one might give also the distances between points of $84$ \% and $50$ \%, and $50$ \% and $16$ \%.
The ratio of these distances gives us the asymmerty of intermixing (shown in Table 1).

\subsection{The computational methods}

  Dynamic TRIM (Transport of Ion in Matter) TRIDYN simulation has been applied to model the ion sputter-removal process (see e.g. ref. \cite{Nastasi}). The input parameters of the code used (TRIDYN \cite{TRIM}), characterizing the pure material, are: atomic number and density. 
($\rho_{Ti} = 56.8$ at/nm$^3$, $\rho_{Pt}=66.2$
at/nm$^3$), bulk binding energies ($0$ for both materials), and the surface binding energies (SBE) of Ti
($4.88$ eV) and Pt ($5.86$ eV).
The simulation provides the atomic concentration along the depth after a given dose of bombarding ions. Having this distribution the Auger intensities of elements present can be calculated using standard equations and parameters like inelastic mean free path (IMFP), backscattering factors, primary current etc  
(the software provided by S.Tougaard is used to calculate the IMFP based on TPP-2M).
Repeating this procedure (simulation of concentration distribution
after additional ion bombardment and calculation of the corresponding Auger intensities)
until the layer sputtered away results in a simulated depth profile, which
might be compared with the measured one. (The complete description of this calculation
 is given in ref. \cite{MM_TRIM}).
As will be seen later the code describes the essence of the experiments, but fails to provide a quantitative agreement. It is not surprizing since we know that this code accounts neither for roughening nor for the recovery processes taking place after the collisional cascade.

 In order to get more insight to the mechanism of interdiffusion 
 classical molecular dynamics simulations have also been used to simulate the ion-solid interaction
(using the PARCAS code \cite{Nordlund_ref}).
Here we only shortly summarize the most important aspects.
A variable timestep
and the Berendsen temperature control is used to maintain the thermal equilibrium of the entire
system. \cite{Allen}. The bottom layers
are held fixed in order to avoid the rotation of the cell.
Periodic boundary conditions are imposed laterarily and a free surface is left for the ion-impacts.
The temperature of the atoms in the
outermost layers was softly scaled towards the desired temperature to provide temperature control and ensure
that the pressure waves emanating from cascades were damped at the borders.
The lateral sides of the cell are used as heat sink (heat bath) to maintain the thermal equilibrium of the entire
system \cite{Allen}.
The detailed description of other technical aspects of the MD simulations are given in \cite{Nordlund_ref,Allen} and details specific to the current system in recent
communications \cite{Sule_PRB05,Sule_NIMB04,Sule_SUCI,Sule_NIMB04_2}.
Atomic collisions have been simulated in a standard way given in refs. \cite{Nordlund_ref,ZBL}.
Recoils have been initialized by giving the kinetic energy of the incoming ion to a lattice atom which is nearby the
impact position.

\begin{table}
\caption[]
{
The experimental and TRIDYN results for depth resolution                            and for the asymmetry of intermixing
for Ti/Pt and Pt/Ti samples.
}
\begin{tabular}{ccccc}
 & exp & TRIDYN & asym. (exp) & asym. (TRYDIN) \\
\hline
 Ti/Pt  & 2.0 & 3.3  & 0.9 & 0.9  \\
 Pt/Ti  & 7.0 & 4.3  & 2.3 & 2.0  \\
\end{tabular}
{\small
The exp and TRIDYN denote our measured and calculated
depth resolutions (nm).
In the 3rd and 4th columns the experimental and the calculated
asymmetry of mixing are given.
}
\end{table}

\begin{figure}[hbtp]
\begin{center}
\caption[]{
The crossectional TEM (XTEM) images of the as received samples.
{\em Upper panel:} Ta/Ti/Pt/Si, the thickness of the Ta, Ti and
Pt layers are $21$, $11$ and $12$ nm, respectively.
{\em Lower panel:} Pt/Ti/Si, the thickness of the Pt and Ti
layers are $13$ and $10$ nm, respectively.
}
\label{fig1}
\end{center}
\end{figure}

  We irradiate the bilayers Pt/Ti and Ti/Pt 
with 0.5 keV Ar$^+$ ions repeatedly with a time interval of 10-20 ps between each of
the ion-impacts at 300 K
which we find
sufficiently long time for the termination of interdiffusion, such
as sputtering induced intermixing (ion-beam mixing) \cite{Sule_NIMB04}.
 The initial velocity direction of the
impacting ion was $10$ degrees with respect to the surface of the crystal (grazing angle of incidence)
to avoid channeling directions and to simulate the conditions applied during ion-sputtering. 
We randomly varied the impact position and the azimuth angle $\phi$ (the direction of the ion-beam).
In order to approach the real sputtering limit a large number of ion irradiation are
employed using automatized simulations conducted subsequently together with analyzing
the history files (movie files) in each irradiation steps.
In this article we present results up to 200 ion irradiation which we find suitable for
comparing with low to medium fluence experiments. 200 ions are randomly distributed
over a $2.0 \times 2.0$ nm$^2$ area which corresponds to $\sim 5 \times 10^{15}$
ion/cm$^2$ ion fluence
and to the removal of few MLs.

 The size of the simulation cell is $11.0 \times 11.0 \times 9.0$ nm$^3$ including
57000 atoms (with 9 monolayers (ML) film/substrate).
At the interface (111) of the fcc crystal is parallel to (0001) of the hcp
crystal
and the close packed directions are parallel.
The interfacial system is a heterophase bicrystal and a composite object of
two different crystals with different
symmetry is created as follows:
the hcp Ti is put by hand on the (111) Pt bulk (and vice versa) and various structures are probed
and are put together randomly. Finally the one which has the smallest
misfit strain prior to the relaxation run is selected.
The difference between the width of the overlayer and the bulk does not exceed $0.2
-0.3$ nm.
The remaining misfit is properly minimized below $\sim 6 \%$ during the relaxation
process so that the Ti and Pt layers keep their original crystal structure and we
 get an
atomically sharp interface.
During the relaxation (equilibration) process the temperature is softly scaled down
to zero.
According to our practice we find that during the temperature scaling down the 
structure
becomes sufficiently relaxed therefore no further check of the structure has been
 done.
Then the careful heating up 
of the system to $300$ K has been carried out. The systems were free from any serious built-in strain
and the lattice mismatch is minimized to the lowest possible level.
The film and the substrate are $\sim 2.0$ and $\sim 6.8$ nm thick, respectively.

 In order to reach the most efficient ion energy deposition at the interface,
we initialize recoils placing the ion above the interface by $1$ nm (and
below the free surface in the 9 ML thick film) at grazing angle of incidence
($10^{\circ}$ to the surface)
with $500$ eV ion energy.
In this way  
we can concentrate directly on the intermixing phenomenon avoiding
many other processes occur at the surface (surface roughening, sputter erosion, ion-induced surface diffusion, cluster ejection, etc.) which weaken energy deposition at the interface.
Further simplification is that channeling recoils are left to leave the cell
and in the next step these energetic and sputtered particles are deleted.

 We used a tight-binding many body potential, developed by Cleri and
Rosato (CR) on the basis of the second moment approximation to the density of states \cite{CR}, to describe interatomic interactions.
This type of a potential gives a good description of lattice vacancies, including atomic migration
properties and a reasonable description of solid surfaces and melting \cite{CR}.
Since the present work is mostly associated with the elastic properties,
melting behaviors, interface and migration energies, we believe the model used should be suitable for this study.
The interatomic interactions are calculated up to the 2nd nearest neighbors
and a cutoff is imposed out of this
region.
This amounts to the maximum interatomic distance of $\sim 0.6$ nm.
 For the crosspotential of Ti and Pt we employ an interpolation scheme \cite{Sule_PRB05,Sule_SUCI,ZBL}
between the respective elements.
The CR elemental potentials and the interpolation scheme for heteronuclear interactions
have widely been used for MD simulations \cite{Sule_PRB05,Stepanyuk2,Goyhenex,Levanov}.
 The Ti-Pt interatomic crosspotential of the Cleri-Rosato \cite{CR} type is fitted to the experimental
 heat of
mixing of the corresponding alloy system \cite{Sule_NIMB04,Sule_NIMB04_2}.
The scaling factor $r_0$ (the heteronuclear first neighbor distance) is calculated as the average of the elemental first neighbor distances.

 The computer animations can be seen in our web page \cite{web}.
Further details  are given in \cite{Nordlund_ref} and details specific to the current system in recent
communications \cite{Sule_PRB05,Sule_NIMB04}.

\section{Results}

\subsection{Experimental results}

 Two typical AES depth profiles recorded on samples of Pt/Ti and Ti/Pt at  eV ion energy,
 are shown, respectively in Fig. 2. It is clear without any detailed evaluation that the Pt/Ti
and Ti/Pt transitions are very different.
While the Ti/Pt transition
is a "normal" one, in the case of Pt/Ti an unusual deep penetration of Pt to the Ti phase is observed.

 According to the experiment a relatively weak intermixing is found in Ti/Pt ($\sigma \approx 2.0$ nm) while an unusually high
interdiffusion occurs in the Pt/Ti bilayer ($\sigma \approx 7.0$ nm).
Moreover we observe a long-range tail in upper Fig 2 for Pt in Pt/Ti while no such tail is found for Ti/Pt
for Pt (neither for Ti, see lower Fig 2).
The depth profiles also show oxygen presence as well. Does the presence of O influence the result? The answer is no which is detailed bellow.

 We have carried out several experiments, using different parts of the sample
and applying various sputtering conditions.
In all cases we recognized oxygen. The oxygen concentration varied considerably.
The oxygen at the metal/substrate interface is due to the native
oxide on the Si substrate.
Since the bulk level of oxygen slightly correlated with the ion current intensity
(the larger
the ion current intensity the lower the oxygen AES signal) part of the oxygen is
contamination occurring during the AES depth profiling process.
On the other hand the interface broadening does not
correlate with the
concentration of
the oxygen
in the Ti/Pt interface.
Thus we conclude
that the atomic transport is not affected by the presence of the slight oxygen contaminant.

 XTEM gives reasonable good estimate of the broadening of the as received sample, which was found to be about $\sim 1,5$ nm (intermixing would result in a gradual change of the contrast; since it was missing we estimate that the initial intermixing is less than $\sim 1,5$ nm). Ion-sputtering might cause surface roughening as well as ion mixing. In case of our experimental conditions (rotated sample, grazing angle of incidence, low relative 
\begin{figure}[hbtp]
\begin{center}
\includegraphics*[height=5.5cm,width=6.6cm,angle=0.]{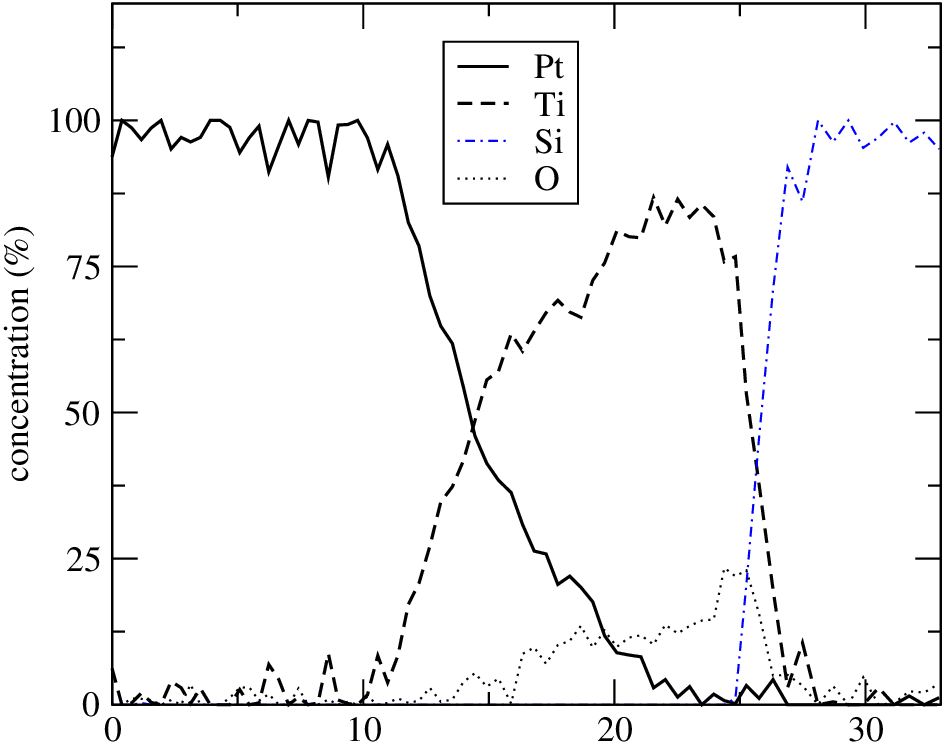}
\includegraphics*[height=5.5cm,width=6.5cm,angle=0.]{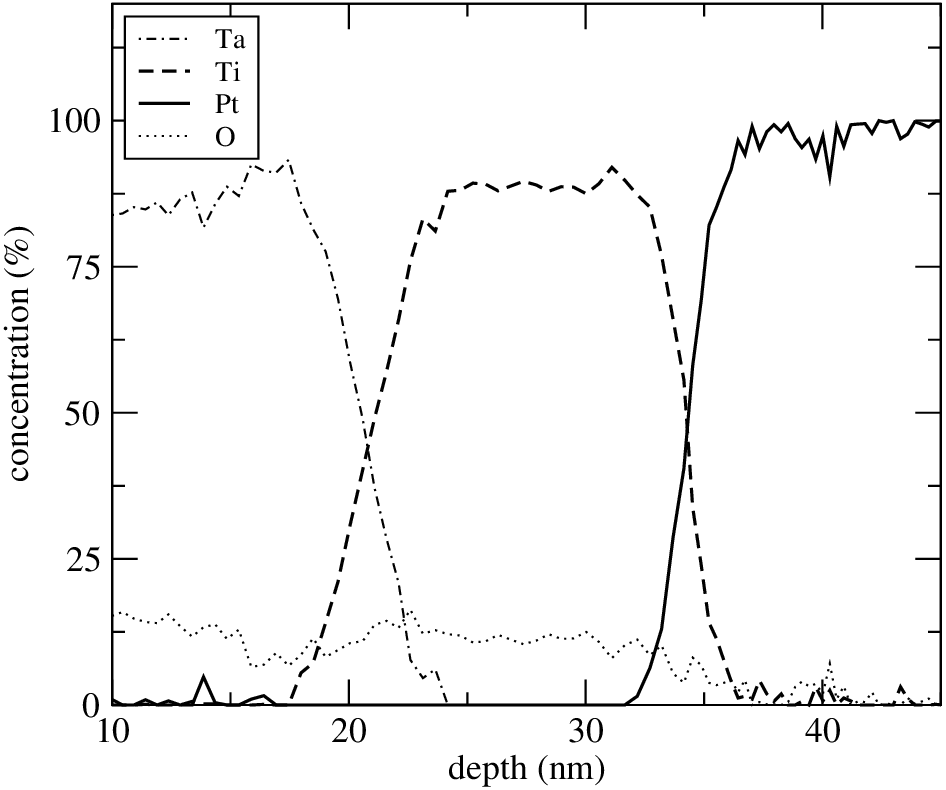}
\caption[]{
The concentration depth profile as a function of the removed layer thickness (nm)
 obtained by AES depth profiling analysis
using ion-sputtering at 500 eV ion energy ($10^{\circ}$ with respect to the surface) in P
t/Ti (upper Fig: 2a) and Ti/Pt (lower Fig: 2b).
}
\label{fig2}
\end{center}
\end{figure}
sputtering yield) the ion bombardment induced roughening is expected to be weak, but cannot be ruled out; thus part of the measured total broadening might be due to roughening.

Unfortunately, using MD simulations it also hard to account for interface roughening, since the wavelength of interface roughening often exceeds the lateral size of the simulation cell. Therefore, we account only for intermixing, and then we estimate the magnitude of roughening as a difference of the measured broadening $\sigma$ and the simulated one.


 The measured depth resolution can be considered as the broadening of the interface $\sigma$ in those cases when the inelastic mean free path of the Auger signal electrons is much smaller than the depth resolution. This condition holds for the case of Ti and Pt.
Moreover, $\sigma$ is defined as $\sigma=\sqrt{\sigma_{ro}^2+\sigma_{mix}^2}$,
where $\sigma_{ro}$ and $\sigma_{mix}$ are the roughening and
intermixing components \cite{Hofmann}.
Unfortunately we can not extract the components even in the as received samples,
hence the $\sigma \approx 1,5$ nm measured by XTEM in the parent samples might include
components both from interface roughening and intermixing.
Also, we can say not too much about the ratio of $\sigma_{ro}$ to $\sigma_{mix}$
after ion-sputtering. We measure an overall value for broadening.
We employ TRIM and molecular dynamics simulations to account for the contribution
of intermixing $\sigma_{mix}$ to broadening $\sigma$.



\subsection{Results obtained by MD and TRIDYN}

  In Table 1 the experimental and TRIDYN results are summarized for the depth resolution
and for the asymmetry of mixing.
It is clear that the TRIDYN describes the essential features of the experiments. It predicts
that broadening is wider when the Pt layer is on the top of the Ti, and that the penetration of
 the Pt into the Ti is strong, while the intermixing of Ti to the Pt phase is weaker.
Considering the absolute values the predicted asymmetry is in excellent agreement with that
of
given by the experiment.
On the other hand the experimental
depth resolution is much
larger for
Pt/Ti than that predicted by TRIDYN.

 It is well known that this code cannot account for the relaxation occurring after the
collisional cascade \cite{ZBL}.
Hence we also carried out MD simulations.

Further advantage of the MD simulations is the applied
many body tight-binding potentials \cite{CR} which are
known to be more accurate and reliable than the two-body potentials used in the TRYDIN code \cite{Nastasi,Sule_NIMB04,ZBL}.

  MD simulations provide $\sigma \approx 8$ ML ($\sim 2.0$ $nm$ and $\sigma \approx 16$ ML ($\sim 4.0$ $nm$ thick interface after $200$ ion impacts, respectively.
The computer animations of the simulations together with the plotted broadening values
at the interface in 
Fig 3 also reveal the stronger
interdiffusion in Pt/Ti \cite{web}.

  Moreover, the applied setup of the simulation cell, in particular the $2.0$ nm film thickness is assumed to be
appropriate for simulating broadening.
Our experience shows that the variation of the film thickness does not affect the
final result significantly, except if ultrathin film is used (e.g. if less than
$\sim 1$ nm thick film). At around $5$ or less ML thick film surface roughening could affect
mixing, and vice versa \cite{Sule_NIMB04_2}.
Also, we do not carry out complete layer-by-layer removal as in the experiment.
It turned out during the simulations that the ions mix the interface the most efficiently
when they are initialized $\sim 1 \pm 0.3$ nm above the interface.
This value is naturally in the range of the projected range of the ions.
Hence, the most of the broadening is coming from this regime of ion-interface distance
.
\begin{figure}[hbtp]
\begin{center}
\includegraphics*[height=6cm,width=7cm,angle=0.]{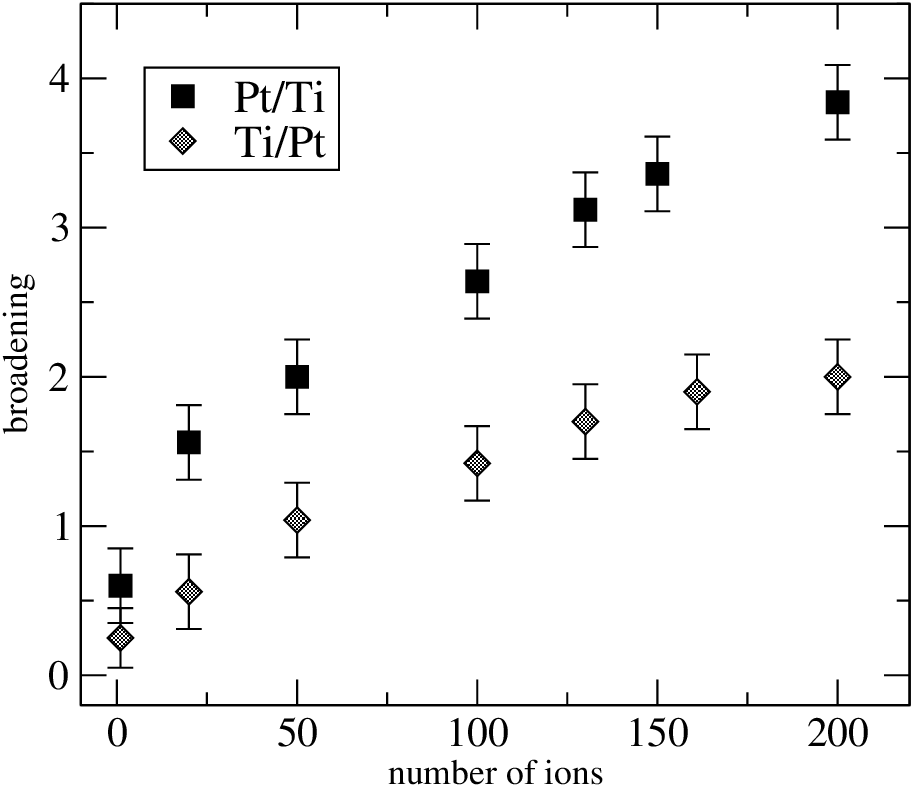}
\caption[]{
The simulated broadening at the interface in nm as a function of
the number of ions at 500 eV ion energy.
The ions are initiated from $1.0$  $nm$ above the interface.
The error bars denote a statistical uncertainty in the measure of broadening.
}
\label{fig3}
\end{center}
\end{figure}
Initializing ions from the surface it takes longer and more ions are needed to
obtain the same level of damage and broadening at the interface as it has been found

by ion-bombarding
from inside the film.

  Indeed, we find that the measured ion-sputtering induced broadening of $\sigma 
\approx 2$ nm for Ti/Pt is in nice agreement with
the simulated value.
Hence we expect that the most of the measured $\sigma$ is coming from intermixing
and interface roughening contributes to $\sigma$ only slightly.
The nice agreement could be due to that the saturation of intermixing (broadening) 
during ion-sputtering is insensitive to
the rate of mixing in the as received samples
($\sigma_0 \approx 1.5$ nm interface width including the interface roughening
in both samples).
This is because
a sharp interface and a weakly mixed one (before bombardment in the as-received samples) lead to the same magnitude of broadening
upon ion-sputtering, because the binary systems reach the same steady state
of saturation under the same conditions (ion energy, impact angle, etc.).
This is rationalized by our finding that
during simulations we start from a sharp interface ($\sigma_0 \approx 0$) and
we get a very similar magnitude of $\sigma$ than by AES.

\section{Discussion}

  Asymmetric AES depth profiles (when the broadening of A/B interface is different from that of B/A) have already been observed \cite{Hofmann,Barna}.
The asymmetric behavior has been explained by the large relative sputtering yield of the elements (preferential sputtering)
\cite{Hofmann,Barna,Berg}.
In the present experiment the relative sputtering yield of
$Y_{Pt}/Y_{Ti} \approx 0.7$ at $500$ eV (at $1500$ eV we find $Y_{Pt}/Y_{Ti} \approx 0.9$ and a similar rate of broadening)
therefore the mechanism is different.
The asymmetry of intermixing is also known during thin film growth
\cite{Buchanan,MM,Jedryka} as well as during concentration gradient driven intermixing processes \cite{Erdelyi}.
Moreover, we found high energy ion-beam mixing results which report
strongly asymmetric interdiffusion e.g. in Au/Cu/Au and in Cu/Au/Cu (with thin tracer impurity layers sandwitched between matrix layers), although these results have not been
discussed in detail \cite{Averback2}.
  However, no reports has been found which present asymmetric IM for ion-sputtering with
vanishing difference of the sputtering yield of the constituents. 
This is interesting result because a new mechanism should be worked out
to explain the preferential interdiffusion of Pt in Pt/Ti.

 The comparison of the measured $\sigma$ with the simulated broadening
using the $84-16$ \% rule in both cases can only be carried out with great care.
In principle, these values are not comparable directly.
However, we make some simple assumptions at this point.
We expect that interface roughening has smaller contribution to broadening
than intermixing.
MD simulations supports this assumption since we find that the
simulated $\sigma$ is in the range of the measured values.
Unfortunately we have no results for the roughening of the samples after ion-bombardment.
However, we expect that if the simulated $\sigma_{mix}$ is comparable with the measured
$\sigma$, than we can expect that the ion-sputtered $\sigma_{ro} \ll \sigma_{mix}$.

  According to the simulations, the mixing of the Pt increases with the fluence,
at $200$ ions irradiation (which is the highest  we reached because of the
limited CPU time) $\sigma \approx 40$ $\hbox{\AA}$ was found. This is
not a saturation value (in contrast to the Ti/Pt case) hence
we can expect further increase in $\sigma$.
The simulated values of broadening are purely coming from intermixing.
We expect that the rest of measured $\sigma$ is coming from roughening.
Nevertheless, MD simulations reproduce the mixing asymmetry and
we are able to capture the essential features of the phenomenon.

 Finally we should also mention that the reason of asymmetry is not fully
understood yet. 
We can rule out the asymmetry of mass effect with the following computer simulation:
if we interchange the atomic masses and leaving all other parameters are unchanged we
also get the asymmetry of mixing. Hence not mass-anisotropy is
responsible for the asymmetry of intermixing.
It could also be that some other parameters, such as e.g. atomic size difference
could explain asymmetric interdiffusion (there is a large difference in atomic
volumes between Ti and Pt, and also the interaction potential of Ti is strongly anharmonic
which could cause the observed anomaly of mixing).
 E.g., it does matter if we ion-bombard the Pt or the Ti film
and the ion-induced injection of Pt to Ti is much easier than that of the Ti atoms to the Pt phase.
This could be due to e.g. the atomic size difference.
However, the verification of this hypothesis goes beyond the scope of the
present paper.

\section{Conclusions}

Performing AES depth profiling on Pt/Ti and Ti/Pt bilayers we have
found strong asymmetry of intermixing depending on the succession of the layers.
 We could reproduce the mixing asymmetry by means of TRIM and MD simulations.
We get a nice agreement for interface broadening in Ti/Pt with experiment while for
Pt/Ti the discrepancy is relatively large.
We conclude from this that interface roughening might has a significant
contribution to broadening in Pt/Ti.
 In Ti/Pt ion-sputtering increases broadening only slightly: in the as received sample
we find $\sigma_0 \approx 1.5$ 
nm,
and $\sigma \approx 2.0$ nm is measured after AES depth profiling.
Although, atomistic simulations reproduce the main features of
interdiffusion,
the mechanism of the asymmetry of intermixing remains, however, unexplained.

{\scriptsize
This work is supported by the OTKA grant F037710
from the Hungarian Academy of Sciences.
We wish to thank to K. Nordlund 
for helpful discussions and constant help.
The work has been performed partly under the project
HPC-EUROPA (RII3-CT-2003-506079) with the support of
the European Community using the supercomputing 
facility at CINECA in Bologna.
The help of the NKFP project of 
3A/071/2004 is also acknowledged.
}

\end{document}